\documentclass[aip,jcp,preprint,dvipdfmx]{revtex4-1}
\usepackage{setspace}
\usepackage{amsmath,amssymb}
\usepackage{mathrsfs}
\usepackage{bm}
\usepackage{color}
\usepackage{graphicx}
\begin{document}


\title{Linear Absorption Spectrum of a Quantum Two-Dimensional Rotator Calculated using a Rotationally Invariant System-Bath Hamiltonian}
\author{Yuki Iwamoto}
\email{iwamoto.y@kuchem.kyoto-u.ac.jp}
\author{Yoshitaka Tanimura}
\email{tanimura.yoshitaka.5w@kyoto-u.jp}
\affiliation{Department of Chemistry, Graduate School of Science, Kyoto University, Kyoto 606-8502, Japan}
\date{\today}

\begin{abstract}
We consider a two-dimensional rigid rotator system coupled to a two-dimensional heat bath. The Caldeira-Leggett (Brownian) model for the rotator and the spin-Boson model have been used to describe such systems, but they do not possess rotational symmetry, they cannot describe the discretized rotational bands in absorption and emission spectra that have been found experimentally. Here, to address this problem, we introduce a rotationally invariant system-bath (RISB) model that is described by two sets of harmonic-oscillator baths independently coupled to the rigid rotator as sine and cosine functions of the rotator angle. Due to a difference in the energy discretization of the total Hamiltonian, the dynamics described by the RISB model differ significantly from those described by the rotational Caldeira-Legget (RCL) model, while both models reduce to the Langevin equation for a rotator in the classical limit. To demonstrate this point, we compute the rotational absorption spectrum defined by the linear response function of a rotator dipole. For this purpose, we derive a quantum master equation for the RISB model in the high-temperature Markovian case. We find that the spectral profiles of the calculated signals exhibit a transition from quantized rotational bands to a single peak after spectrum collapse. This is a significant finding, because previous approaches cannot describe such phenomena in a unified manner.
\end{abstract}


\maketitle

\section{Introduction}\label{sec:intro}
In order to understand molecular dynamics, rotational motion is as important as translational and vibrational motion.\cite{Burstein} Recent theoretical and experimental works have demonstrated the importance of the interplay between the quantum nature of a system and environmental noise. While the quantum properties of translational and vibrational relaxation processes have been thoroughly investigated using spin-Boson and Brownian models,\cite{GrabertPR88,Weiss08, Breuer, Kuhn, Nitzan,Tanimura-2006} the study of rotational relaxation has been limited due to the lack of a simple rotational system-bath model that can be solved numerically. In infrared (IR), far-IR, rotational-Raman, and dielectric absorption and dispersion spectroscopy studies, the quantum nature of rotational relaxation is characterized by equally spaced peaks, called rotational bands, that arise from energy transitions among the quantized rotational states of molecules. From gas phase experiments, it is known that these rotational peaks merge into a single broadened peak at the center of the rotational bands when the gas pressure or density becomes sufficiently high. Then, the width of this merged peak progressively narrows as the gas pressure or density increases.

Such phenomena have been described in terms of adiabatic and nonadiabatic collisions between rotational molecules and gas molecules by introducing semi-empirical relaxation terms in the quantum Liouville equation on the basis of scattering theory.\cite{Burstein, Burstein-1974} Various extensions of such approaches have been used to derive relaxation terms in the investigation of rotational spectra.\cite{Burstein-1989, Burstein-1992, Filippov-1995, Coffey} {Examples involve a quantum J-diffusion model, which describe quantum nature of IR spectra, while it possess a correct classical limit in the underdamped and overdamped cases.\cite{McClung-1974, Gelin-1999}
However, in such systems, the mechanism of the relaxation process is not clear, due to its phenomenological nature.} In addition, the effect of resulting from the quantum nature of heat bath, in particular in the non-Markovian case, are not clear.

In this paper, we describe these phenomenon on the basis of a system-bath model approach developed in the context of open quantum dynamics theories. In such a treatment, a rotationally invariant system-bath Hamiltonian, satisfying $H(\theta)=H(\theta+2 \pi)$, where $\theta$ is the rotator angle, must be employed to study dynamics in order to avoid unphysical behavior.\cite{Curruthers} 
While the classical description of a Brownian rotator, whose dynamics are equivalent to Langevin dynamics, is appropriate for describing classical rotational relaxation, the quantum description, which has been studied using the Caldeira-Leggett model,\cite{Caldeira-AP-1983} does not exhibit rotational bands.\cite{Suzuki-2001,Suzuki-2002}  This is because in the rotational Caldeira-Leggett (RCL) model,  the total system does not possess rotational symmetry. It should be noted, however, that in this model, the rotational invariance of the rotator itself is recovered after tracing over the bath degrees of freedom, and analytically exact expressions for the linear and nonlinear response functions has been obtained.\cite{Suzuki-2002, Suzuki-2003} 

As an extension of the standard Brownian model, a periodic system-bath (PSB) model has been used in studies of inelastic nuclear scattering (NIS) and nuclear magnetic resonance (NMR).\cite{Wurger-1989-1, Wurger-1989-2,Wurger-1990,Braun-1994, Szymanski-1999, Szymanski-2005, Szymanski-2015, Szymanski-2017} This approach assumes that the system-bath interaction $H_I = V(\theta)X(t) $ satisfies $V(\theta) = V(\theta + 2 \pi/N)$ for a $C_N$ symmetric rotator, where $V(\theta)$ is the system side of the system-bath interaction and $X(t)$ is the collective coordinate of the bath, which corresponds to noise.
While the quantum master equation derived from the PSB model can describe rotational bands, the overdamped peak predicted by this model is different from that which arises from the spectral collapse peak predicted by the classical Langevin approach. 

In order to overcome this limitation, here we introduce a rotationally invariant system-bath (RISB) model described by a rotationally invariant system-bath Hamiltonian. This Hamiltonian consists of a two-dimensional rigid rotator independently coupled to the $x$ and $y$ elements of a two-dimensional harmonic oscillator bath with sine and cosine functions of the rotator angle $\theta$ as $H_I = \cos(\theta) X(t) + \sin(\theta) Y(t)$, where $X(t)$ and  $Y(t)$ are the collective coordinates of the baths in the $x$ and $y$ directions, respectively. This Hamiltonian was introduced by Gefen, Ben-Jacob and Caldeira in order to study a dissipative driven system, specifically, a current-biased tunnel junction.\cite{Caldeira-1987}
We found that this model is also suitable for the description of rotational spectra, because the model is rotationally invariant, and because the equation of motion described by this Hamiltonian reduces to the Langevin equation in the overdamped limit.  Moreover, it is possible to extend this model so that it can describe {the effects of anisotropic environments, such as anisotropic 2D crystals.}

In order to demonstrate some of the above-mentioned features, we derive a Markovian master equation without imposing the rotating wave approximation (RWA) for the RISB model that is realized when we assume an Ohmic spectral distribution for a high temperature bath. In the overdamped case, this equation reduces to the Fokker-Planck (or Kramers) equation, which is equivalent to the Langevin equation.
With this model, we can describe the rotational spectrum from the quantum regime to the classical overdamped regime uniformly as a function of the system-bath coupling and bath temperature. 

This paper is organized as follows.
In Sec. II, we describe the model and discuss its theoretical foundation.
In Sec. III, we introduce the linear response function for the absorption spectrum of the rotator.
In Sec. IV, we present numerical results and discussion.
Section V is devoted to concluding remarks.

\section{The Model and Its Theoretical foundation}
\subsection{A rotationally invariant system-bath model}
We consider a two-dimensional rigid rotator system described by
\begin{align}
\hat{H}_S &= \frac{\hat{L}^2 }{2 I} + U(\hat{\theta}), 
\label{eq:RFR}
\end{align}
where $\hat{L}$, $\hat{\theta} $ and $I$ are the angular momentum, angular coordinate and moment of inertia of the rigid rotator, and $U(\hat{\theta})$ is a periodic potential that satisfies $U(\hat{\theta})= U(\hat{\theta}+2\pi)$.
{Examples of two-dimensional rotator systems include the rotational motion of benzene about the C$_6$ axis and the methyl group rotation of toluene.}


The rotator system is independently coupled to two heat baths in the $x$ and $y$ directions (a two-dimensional heat bath) through sine and cosine functions of $\theta$. The total Hamiltonian is then given by
\begin{align}
\hat{H}_{tot} &= \hat{H}_S + \hat{H}_{I+B}^{x} + \hat{H}_{I+B}^{y},
\label{eq:tot_H}
\end{align}
where  
\begin{align}
\hat{H}_{I+B}^{\alpha} = \sum_k \left\{
\frac{(\hat{p}_{k}^{\alpha})^2}{2 m_k^{\alpha}} + \frac{1}{2}m_k^{\alpha}  (\omega_k^{\alpha})^2 \left(\hat{q}_k^{\alpha}  -
\frac{c_{k}^{\alpha} \hat V^{\alpha}}{m_{k}^{\alpha} (\omega_{k}^{\alpha})^{2}} \right)^2\right\},
\label{eq:Balpha}
\end{align}
and $m_k^{\alpha}$, $\hat{p}_k^{\alpha}$, $\hat q_k^{\alpha}$ and $\omega_k^{\alpha}$ are the mass, momentum, position and frequency variables of the $k$th bath oscillator mode in the $\alpha = x$ or $y$ direction. Here, we set $\hat{V}^x = \cos\hat{\theta}$ and $\hat{V}^y = \sin\hat{\theta}$, and $c_k^{\alpha}$ is the system-bath coupling constant. From Eqs.\eqref{eq:RFR}-\eqref{eq:Balpha}, it is seen that the two terms in the interaction part of the Hamiltonian are assumed to take the forms $\hat{H}_{I}^x = -\cos(\hat{\theta})\hat{X}$ and $\hat{H}_{I}^y = -\sin(\hat{\theta})\hat{Y}$, where $\hat{X}\equiv \sum_k c_k^x \hat{q}_k^x$ and $\hat{Y}\equiv \sum_k c_k^y \hat{q}_k^y$ are the interaction coordinates in the $x$ and $y$ directions.
Note that we have introduced the counter terms $\sum_k (c_k^x)^2 \cos^2(\hat{\theta})/2m_k^x (\omega_k^x)^2$ and $\sum_k (c_k^y)^2 \sin^2(\hat{\theta})/2m_k^y (\omega_k^y)^2$ to maintain the translational symmetry of the Hamiltonian in the $x$ and $y$ directions. We can regard these baths to arise from, for example, the $x$ and $y$ components of the local electric field due to the surrounding molecules. In the case of an electric molecular dipole, the interaction between the rotator and the environments is described by $\hat{H}_{I}^x \propto \cos(\hat{\theta})X(t)$ and $\hat{H}_{I}^y \propto \ \sin(\hat{\theta})Y(t)$, where $X(t)$ and $Y(t)$ are the components of the local electric field arising from the fluctuations of the surroundings molecules. The harmonic baths are characterized by spectral density functions defined as
\begin{align}
J^{\alpha}(\omega) = \frac{\pi}{2} \sum_k \frac{(c_k^{\alpha})^2}{m_k^{\alpha} \omega_k^{\alpha}} \delta(\omega - \omega_k^{\alpha}),
\label{Jomega} 
\end{align}
where $\alpha$ represents $x$ or $y$. It should be noted that $J^{x}(\omega)$ and $J^{y}(\omega)$ need not be the same. In particular, they will differ when the surrounding environment is anisotropic.

With the above Hamiltonian, the system dynamics can be derived numerically rigorously in the case of non-Markovian noise using the hierarchal equations of motion (HEOM) approach.\cite{Tanimura-1989, Tanimura-1990, Ishizaki-2005, Tanimura-2006, Tanimura-2014,Tanimura-2015} However, in the case of the multiple heat baths, the HEOM approach is extremely computationally demanding. For this reason, here we restrict our analysis to the simple Markovian case.

\subsection{Quantum master equation for the RISB model}
In the case of a weak system-bath coupling, the generalized master equation approach is appropriate for the study of quantum dissipative dynamics,\cite{Breuer, Kuhn, Nitzan} while this equation exhibits pathological behavior in the strong coupling case.\cite{Tanimura-2014,Tanimura-2015} Without employing the RWA, the generalized master equation for the reduced density matrix of the system, $\hat{\rho}(t)$, in the Schr{\"o}dinger representation derived from Eqs. \eqref{eq:RFR}-\eqref{eq:Balpha} is expressed as\cite{Kuhn}
\begin{align}
\frac{\partial}{\partial t} \hat{\rho}(t)  &= -\frac{i}{\hbar}[\hat{H}_S' , \hat{\rho}(t)] - \frac{1}{\hbar^2} \int^{t \textcolor{blue}{- t_0}}_{0} d \tau \left(\hat \Gamma^x(\tau) \hat{\rho}(t - \tau) + \hat \Gamma^y(\tau) \hat{\rho}(t - \tau)\right),
\label{eq:GQME}
\end{align}
where $\hat{H}_S'$ is the system Hamiltonian with the counter terms, and
\begin{align}
\hat \Gamma^{\alpha}(\tau) \hat{\rho}(t - \tau) &\equiv C^{\alpha} (\tau) [\hat{V}^{\alpha} , \hat{G}_S(\tau) \hat{V}^{\alpha} \hat{\rho}(t - \tau) \hat{G}_S^{\dagger}(\tau)] \notag \\
& - C^{\alpha} (-\tau)[ \hat{V}^{\alpha} ,  \hat{G}_S(\tau) \hat{\rho}(t - \tau) \hat{V}^{\alpha} \hat{G}_S^{\dagger}(\tau)]
\label{eq:Gamma}
\end{align}
is the damping operator, in which
\begin{align}
C^{\alpha} (\tau) &= \hbar \int^{\infty}_0 \frac{d \omega}{\pi} J^{\alpha} (\omega) \left[ \coth \left(\frac{\beta \hbar \omega}{2}\right) \cos(\omega \tau) - i \sin(\omega \tau)\right]
\label{eq:spectral}
\end{align}
is the bath correlation function for the $\alpha=x$ and $y$ baths. Here, $\beta= 1/k_B T$ is the inverse temperature of the environments divided by the Boltzmann constant, $k_B$, and $\hat{G}_S(\tau)$ is the time evolution operator of the system. We assume the Markovian case described by an Ohmic spectral distribution $J^{\alpha}(\omega) = \eta^{\alpha} \omega$, where $\eta^{\alpha}$ is the friction coefficient, in the high temperature case, in which we have $\coth(\beta\hbar\omega/2)\approx 2/(\beta\hbar\omega)$. The bath correlation function is then expressed as
\begin{align}
C^{\alpha} (\tau) = \eta^{\alpha} \left( \frac{2 }{\beta}  + i
 \hbar  \frac{d}{d\tau} \right) \delta(\tau),
\end{align}
and we have $\int^{t \textcolor{blue}{-t_0}}_{0} d \tau \hat \Gamma^{\alpha}(\tau) \hat{\rho}(t - \tau) = \hat {\bar \Gamma}^{\alpha} \hat{\rho}(t)+{i \hbar \eta^{\alpha} }\delta(0) [(\hat{V}^{\alpha})^2, \hat{\rho}(t)]$. The imaginary term on the right-hand side (RHS) is canceled by the counter terms. Thus we have
\begin{align}
\frac{\partial}{\partial t} \hat{\rho}(t)  &= -\frac{i}{\hbar}[\hat{H}_S , \hat{\rho}(t)] - \frac{1}{\hbar^2}  \hat {\bar \Gamma}^x(\tau) \hat{\rho}(t) -\frac{1}{\hbar^2}\hat {\bar \Gamma}^y \hat{\rho}(t),
 \label{eq:QME}
\end{align}
where 
\begin{align}
\hat {\bar \Gamma}^{\alpha}  \hat{\rho}(t)&= 
 \frac{\eta^{\alpha}}{\beta}  \left(  [\hat{V}^{\alpha},  \hat{V}^{\alpha} \hat{\rho}(t) ] - [ \hat{V}^{\alpha} ,   \hat{\rho}(t ) \hat{V}^{\alpha}]  
\right) +\frac{i\hbar \eta^{\alpha}}{2 } \left[(\hat{V}^{\alpha})^2 ,  \frac{d \hat{\rho}(t - \tau)}{d \tau}|_{\tau = 0} \right]\notag \\
&- \frac{\eta^{\alpha}}{2 } \left([\hat{V}^{\alpha} ,  \hat{H}_S \hat{V}^{\alpha} \hat{\rho}(t ) ] + [ \hat{V}^{\alpha} ,   \hat{H}_S \hat{\rho}(t ) \hat{V}^{\alpha} ] - [\hat{V}^{\alpha} ,  \hat{V}^{\alpha} \hat{\rho}(t ) \hat{H}_S] - [ \hat{V}^{\alpha} ,   \hat{\rho}(t ) \hat{V}^{\alpha} \hat{H}_S]  \right).
\label{eq:anisoQME}
\end{align}
For an isotropic environment, with $\eta=\eta^{x}=\eta^{y}$, the second term on the RHS of Eq. \eqref{eq:anisoQME} vanishes, because we have the relation $(\hat{V}^x)^2 + (\hat{V}^y)^2 = const.$   Thus, as the quantum master equation (QME) for the RISB model in the isotropic case, we obtain
\begin{align}
\frac{\partial}{\partial t} \hat{\rho}(t)  &= -\frac{i}{\hbar}[\hat{H}_S , \hat{\rho}(t)] -\frac{\eta}{\beta \hbar^2}  \left(  [\hat{V}^{x},  \hat{V}^{x} \hat{\rho}(t) ] - [ \hat{V}^{x} ,   \hat{\rho}(t ) \hat{V}^{x}]  
\right) \notag \\
& -\frac{\eta}{\beta \hbar^2}  \left(  [\hat{V}^{y},  \hat{V}^{y} \hat{\rho}(t) ] - [ \hat{V}^{y} ,   \hat{\rho}(t ) \hat{V}^{y}]  
\right)\notag \\
&+ \frac{\eta}{2 \hbar^2} \left([\hat{V}^{x} ,  \hat{H}_S \hat{V}^{x} \hat{\rho}(t ) ] + [ \hat{V}^{x} ,   \hat{H}_S \hat{\rho}(t ) \hat{V}^{x} ] - [\hat{V}^{x} ,  \hat{V}^{x} \hat{\rho}(t ) \hat{H}_S] - [ \hat{V}^{x} ,   \hat{\rho}(t ) \hat{V}^{x} \hat{H}_S]  \right)  
 \notag \\
&+ \frac{\eta}{2 \hbar^2} \left([\hat{V}^{y} ,  \hat{H}_S \hat{V}^{y} \hat{\rho}(t ) ] + [ \hat{V}^{y} ,   \hat{H}_S \hat{\rho}(t ) \hat{V}^{y} ] - [\hat{V}^{y} ,  \hat{V}^{y} \hat{\rho}(t ) \hat{H}_S] - [ \hat{V}^{y} ,   \hat{\rho}(t ) \hat{V}^{y} \hat{H}_S]  \right).
\label{eq:QMEiso}
\end{align}
Note that, in this isotropic case, Eq. \eqref{eq:QMEiso} holds either with or without the counter terms, because the second term on the RHS of Eq. \eqref{eq:anisoQME} is canceled in either case, due to the 
relation $(\hat{V}^x)^2 + (\hat{V}^y)^2 = const.$  Moreover, this equation is invariant under rotational motion, $\hat{\theta} \to \hat{\theta} + \alpha$, because the relaxation operators of this equation possess rotational invariance.

For numerical calculations, an eigenstate representation of the QME is more useful than the angular coordinate representation. In the case of a free rotator, i.e. when $U(\theta)=0$, the above equation can be expressed as
\begin{align}
\frac{\partial}{\partial t} \rho_{a,b}(t)  &= -i \omega_0 ( a^2 - b^2 )\rho_{a,b} - \frac{ \eta}{\beta \hbar^2} \left(2 \rho_{a,b} - \rho_{a+1,b+1} - \rho_{a-1,b-1}  \right) \notag \\
&\quad + \frac{\eta {\omega_0} }{2 \hbar} \left( (a + b + 1) \rho_{a+1,b+1} - ( a + b - 1) \rho_{a-1,b-1} + 2 \rho_{a,b}  \right),
\label{eq:RRE}
\end{align}
where $\rho_{a,b} \equiv \langle a | \hat{\rho} | b \rangle \notag$ for the eigenstates $|a\rangle$ and $|b\rangle$ with eigenvalue $a$ and $b$ (satisfying $- \infty < a, b < \infty$) and $\omega_0 \equiv \hbar/ 2I$.

\subsection{Fokker-Planck Equation for the RISB model}
In the case of a strong system-bath coupling, the rotational motion relaxes quickly due to the large viscosity of the environment.
Thus, in this case, the periodic nature of the rotator can be ignored, and for this reason, the domain of $\theta$ can be extended from $- \pi \le \theta < \pi$ to $- \infty < \theta < \infty$. Then, Eq.\eqref{eq:QMEiso} in the case $U(\theta)=0$ becomes
\begin{align}
\frac{\partial}{\partial t} \rho(\theta,\theta',t) &= \left(  \frac{i \hbar}{2 I} \frac{\partial^2}{\partial \theta^2} - \frac{i \hbar}{2 I} \frac{\partial^2}{\partial \theta'^2} \right. \notag \\
&\quad \left. - \frac{\eta}{\beta \hbar^2 } \left \{ [ \sin( \theta) - \sin( \theta') ]^2 + [\cos( \theta) - \cos( \theta') ]^2      \right\} \right. \notag \\
&\quad   - \frac{\eta}{2 I} \sin (\theta - \theta')\left (\frac{\partial}{\partial \theta} - \frac{\partial}{\partial \theta'} \right) \notag \\
&\quad \left.+ \frac{\eta}{2 I} \left \{ [ \sin( \theta) - \sin( \theta') ]^2 + [\cos( \theta) - \cos( \theta') ]^2      \right\}   \right)\rho(\theta,\theta',t).
\label{eq:RFPE}
\end{align}
In the Wigner representation,\cite{Wigner-1984}
the above equation further reduces to the quantum Fokker-Planck equation (QFPE), expressed as\cite{CaldeiraPhysica83,Tanimura-2006}(see Appendix A)
\begin{align}
\frac{\partial}{\partial t} W(p,\theta,t) &= -\frac{p}{I} \frac{\partial}{\partial \theta}W(p,\theta,t) + \frac{\eta}{I} \frac{\partial }{\partial p}\left( p+ \frac{I}{\beta} \frac{\partial}{\partial p} \right) W(p   , \theta , t).
\label{FPEQ}
\end{align}
Note that in the QFPE approach, it is possible to include the contribution of the potential term by introducing the Wigner representation of  $U(\theta)$.\cite{Frensly, TanimuraJCP92, SakuraiJPC11, KatoJPCB13}
In the present case, the QFPE and the classical Fokker-Planck equation (CFPE) are identical, because we do not have a potential term.  As we show below, while the CFPE can be applied in the weak coupling case at low temperature, the QFPE can be applied only in the overdamped case at high temperature, because we employed the high temperature assumption in deriving Eq.\eqref{FPEQ}, in addition to extending the domain of $\theta$ to $- \infty < \theta < \infty$.

\subsection{Classical Langevin Equation}
In the classical case, we can derive the classical Langevin equation (CLE) form the Hamiltonian given in Eqs. \eqref{eq:RFR}-\eqref{eq:Balpha}. From the Hamilton canonical equations, we obtain the following set of differential equations:
\begin{align}
I \ddot{\theta} +U(\theta)&=  - \eta^x \sin\theta \sum_k c_k x_k + \eta^y \cos\theta \sum_k  c_k y_k, 
\end{align}
and
\begin{align}
m_k \ddot{x}_k &= -m_k \omega_k^2 x_k + c_k \cos\theta, \\
m_k \ddot{y}_k &= -m_k \omega_k^2 y_k + c_k \sin\theta \notag.
\end{align}
Then, after eliminating $x_k$ and $y_k$, we obtain the generalized Langevin equation as
\begin{align}
I \ddot{\theta}  +  \int^t_0 dt' \left[\gamma^x(t-t') \sin\theta(t) \sin\theta(t') + \gamma^y(t-t')\cos\theta(t) \cos\theta(t')) \right] \dot{\theta}(t') \notag \\
= - \sin \theta(t) \zeta^x(t)+ \cos \theta(t)\zeta^y(t),
\end{align}
where $\gamma^{\alpha} (t)$ and $\zeta^{\alpha}(t)$ represent the frictional and random forces in the $\alpha$ direction, defined as
\begin{align}
\gamma^{\alpha}(t-t') 
\equiv \frac{2}{\pi} \int^{\infty}_0 d \omega \frac{J^{\alpha}(\omega)}{\omega} \cos(\omega (t-t')), 
\end{align}
and
\begin{align}
\zeta^{\alpha}(t) &\equiv - \sum_k c_k^{\alpha} \left\{ \frac{p_{k}^{\alpha} (0)}{m_k^{\alpha} \omega_k^{\alpha}} \sin(\omega_k^{\alpha} t) + (x_k^{\alpha} (0) -\frac{c_k^{\alpha}}{m_k^{\alpha} (\omega_k^{\alpha})^2} \cos(\theta(0)) ) \cos(\omega_k^{\alpha} t) \right\}.
\end{align}

For the isotropic case, with $\langle \zeta^x (t) \zeta^x(t')\rangle=\langle \zeta^y(t) \zeta^y(t')\rangle $, we have
\begin{align}
\left \langle \zeta(t) \zeta(t') \right \rangle
&= \frac{1}{\beta}  \sum_k \frac{c_k^2}{m_k \omega_k^2} \cos \omega_k (t-t') \cos(\theta(t) - \theta(t')) \notag \\
&= \frac{1}{\beta}  \cos(\theta(t) - \theta(t')) \gamma(t-t'),
\end{align}
where we have set the mean value of $\zeta^{\alpha}(t)$ to 0, i.e. $\langle \zeta^{\alpha} (t) \rangle = 0$. In the Ohmic case, with $J(\omega) = \eta \omega$, the friction kernel is given by
$\gamma(t-t') = 2 \eta \delta(t-t')$.
Then, the CLE is derived as
\begin{align}
I \ddot{\theta} + \eta \dot{\theta}  +U(\theta)&= \zeta(t), 
\label{eq:Langevin1}
\end{align}
with
\begin{align}\langle \zeta(t) \zeta(t') \rangle &= \frac{2 \eta}{\beta} \delta(t-t').
\label{eq:Langevin2}
\end{align}
In the case $U(\theta)=0$, the dynamics described by the CLE are equivalent to those described by the CFPE, which are identical to those described by Eq.\eqref{FPEQ}. However, although the QFPE presented in Eq.\eqref{FPEQ} is valid only in the overdamped case at high temperature, there is no such limitation on the CFPE.

\section{Linear Absorption Spectrum}
\subsection{Response function}

The linear absorption spectrum of a molecular dipole moment $\hat \mu = \mu_0 \cos \theta$ is expressed as\cite{Mukamel}
\begin{align}
\sigma[\omega] = \mathrm{Im} \left[\mu_0^2 \int^{\infty}_0 dt e^{i \omega t} R(t) \right],
\end{align}
where $R(t)$ is the response function defined as
\begin{align}
R(t) \equiv \frac{i}{\hbar} \langle [\cos(\theta(t)) , \cos(\theta(0))] \rangle.
\end{align}
In order to calculate $R(t)$ using an equation of motion approach, we express the response function as $R(t)= \frac{i}{\hbar} \mathrm{Tr} \left\{ \hat{\mu} \mathscr{G}(t)\hat{\mu}^{\times} \hat{\rho}^{eq} \right\},$ where the hyperoperator $^{\times}$ is defined as $\hat{A}^{\times} \hat{B} \equiv [\hat{A} , \hat{B} ]$, and $\mathscr{G}(t)$ is the Green function of the system Hamiltonian without a laser interaction.\cite{Tanimura-2006} In the reduced equation of motion approach, the density matrix is replaced by a reduced one, and the Liouvillian in $\mathscr{G}(t)$ is replaced using the QME. Then we evaluate the absorption spectrum in the following steps. (i) The system is initially in the equilibrium state: $\hat{\rho}_{eq} = e^{- \beta \hat{H}_S}$. (ii) The system is excited by the first interaction $\hat{\mu}^{\times}$ at $t=0$. (iii) The time evolution of the perturbed elements is then computed by integrating Eq.\eqref{eq:RRE} using the fourth-order Runge-Kutta method. (v) $R(t)$ is calculated from the expectation value of $\hat{\mu}$. Finally, performing a fast Fourier transform, we obtain $\sigma(\omega)$.

\subsection{Kubo Oscillator}

In the classical case, we calculate the correlation function defined as 
\begin{align}
C(t) &\equiv  \langle \cos(\theta(0)) \cos(\theta(t)) \rangle_{cl},
\end{align}
where $\langle  \rangle_{cl}$ represents the thermal average over the classical distribution.  We can obtain the response function from $C(t)$ using the fluctuation-dissipation theorem in the classical case, expressed as $R[\omega]=i\omega C[\omega]/\beta$, where $R[\omega]$ and $C[\omega]$ are the Fourier transforms, of $R(t)$ and $C(t)$, respectively. \cite{Kubo}  Then, the rotational spectrum in the classical case is expressed as
\begin{align}
\sigma[\omega]= \frac{i \omega \mu_0^2 }{\beta} C[\omega].
\end{align}
This function is analytically calculated from the CLE given in Eqs.\eqref{eq:Langevin1} and \eqref{eq:Langevin2} as follows.\cite{McConnel} First, we consider the rotational matrix
\begin{align}
\bm{R}(t) = \begin{pmatrix}
\cos\theta(t) & -\sin\theta(t) \\
\sin\theta(t) &\cos\theta(t) 
\end{pmatrix}.
\end{align}
The time derivative of $\bm{R}(t)$  is given by
\begin{align}
\frac{d \bm{R}(t)}{dt} &=
\begin{pmatrix}
0 & 1 \\
-1 & 0
\end{pmatrix}
\bm{R}(t) \omega(t),
\end{align}
where $\omega = \dot{\theta}$ is the angular frequency. In the Kubo oscillator model, this angular frequency is regarded as a stochastic variable. \cite{Kubo-1954, Anderson}  Here, we consider the case in which  $\omega$ is governed by the CLE, Eqs.\eqref{eq:Langevin1} and \eqref{eq:Langevin2}. 
For this reason, its correlation function is given by $\langle \omega(0) \omega(t) \rangle =  e^{-\frac{\eta}{I} t}/I \beta$.  Furthermore, because $\omega$ is governed by the Langevin equation, it is a Gaussian stochastic variable, and hence we have $\langle \bm{R}(t) \rangle = \exp \left[ -\gamma (\frac{\eta}{I} t - 1 + e^{- \frac{\eta}{I} t}) \right] \bm{I}$ for the initial condition $\theta(0) = 0$, where $\gamma = I/\beta \eta^2$ and $\bm{I}$ is the two-dimensional unit matrix.\cite{McConnel}
The unit vector in the direction of the rigid rotator is denoted by $\bm{n}(t)$. The correlation function of $\bm{n}(t)$ is given by $\langle \bm{n}(0) \cdot \bm{n}(t) \rangle =\langle \bm{R}(t) \rangle.$ Thus we have $\langle \bm{n}(0) \cdot \bm{n}(t) \rangle = \langle \cos(\theta(t)) \rangle = \langle \cos(\theta(0)) \cos(\theta(t)) \rangle$ for $\theta(0) = 0$, so that
\begin{align}
C(t)= \exp \left[ -\gamma \left(\frac{\eta}{I} t - 1 + e^{- \frac{\eta}{I} t}\right) \right].
\end{align}

Thus, in the weak damping regime, the response function becomes a Gaussian-like profile, and we have
\begin{align}
\sigma[\omega]= \frac{\mu_0^2\eta}{I \sqrt{2 \pi \gamma}}  \omega \exp \left( -\frac{\eta^2 \omega^2}{2 \gamma I^2} \right),
\label{eq:Gaussian}
\end{align}
while in the strong damping regime, it becomes a Lorentzian-like profile, and we have
\begin{align}
\sigma[\omega]= \frac{\mu_0^2 e^{\delta}}{\pi} \frac{\gamma^2 \omega}{\omega^2 + \gamma^2},
\label{eq:Lorentzian}
\end{align}
where $\delta$ is a small real constant from the phase.
This change in profile from a Gaussian to a Lorentzian is known as a motional narrowing in the context of NMR\cite{Kubo}. This is regarded as a classical phenomenon, as this derivation suggests.

\section{Results and Discussion}
In what follows, we study absorption spectra for the RISB and RCL models. 
We consider (a) the moderate-temperature case (with $\beta \hbar \omega_0 = 0.2$) and (b) the high-temperature case (with $\beta \hbar \omega_0 = 0.02 $).
Although we assumed the high temperature limit to derive the QME in the RISB case, this condition is easily satisfied for measurements in molecular rotational spectroscopy experiments. For example, for the rotational motion of a methyl group, the moment of inertia is $2.1 \times 10^{-47} kg \cdot m^2$, and we have $\beta \hbar \omega_0 = 0.05 \ll 1$ at room temperature. In the following, we set $\mu_0=1$. In the RCL case, we employed an analytically exact solution for the absorption spectrum that is obtained using the path integral approach.\cite{Suzuki-2001,Suzuki-2002, Suzuki-2003}

In Fig. 1, we plot rotational absorption spectra calculated from the RISB model, the RCL model, and the CLE given in Eqs. \eqref{eq:Langevin1} and Eq. \eqref{eq:Langevin2}, which corresponds to the classical limit of the RISB and RCL models for various values of the coupling strength, $\bar{\eta} = \eta / \hbar$, in (a) the moderate case and (b) the high temperature case. The spectra in the CL case were calculated from the analytical expression presented in Refs. \onlinecite{Suzuki-2002, Suzuki-2003}.

Figure 1(a-i)-(a-iv) illustrate the rotational absorption spectra for the moderate temperature case. First, it should be noted that, although the quantum RCL results are slightly lower than the Langevin results, the overall profiles are very similar, because the quantum effects are minor in the RCL model in this temperature regime. In the very weak coupling case, depicted in Fig. 1(a-i), the quantum RISB results exhibit discretized rotational bands arising from quantum transitions $J \to J \pm 1$ with energy differences $E_{J+1} - E_{J} = \omega_0 (2 J + 1)$, while the quantum RCL results are similar to the classical results. The existence of these rotational bands is due to the fact that the total Hamiltonian of the RISB model possesses rotational symmetry. Contrastingly, the RCL model possesses rotational symmetry only for the system part. In the quantum RISB case, the profiles of the absorption peaks are determined from the differential equation for $\partial \rho_{J,J + 1}(t)/\partial t$ given in Eq. \eqref{eq:RRE}: In the weak coupling case, we can ignore the contribution from the RHS terms with $\rho_{J+1,J + 2}$  and $\rho_{J-1,J}$, and as a result, we have $\rho_{J, J+  1}(t) = e^{- \eta (1- 2/\beta)t + i(2J+ 1)t}$. Thus, the peak profile in the RISB case is expressed as a sum of Lorentzian functions, $\sigma(\omega) \propto \sum_J {\eta(1- 2/\beta)}/[{(\omega - 2J - 1)^2 + \eta^2(1- 2/\beta)^2}]$, and the width of each peak is given by $\eta (1- 2/\beta)$. Contrastingly, we observe a broadened peak only in the quantum CL cases, as in the classical case, because the quantum CL model does not possess rotational symmetry and the transition energy of rotational motion become continuous.

When the system-bath coupling becomes slightly larger, as in the case of Fig. 1(a-ii), the contribution from the other terms with $\rho_{J+1,J + 2}$  and $\rho_{J-1,J}$ plays a significant role. As a result, in this case, the spectral profiles deviate from the Lorentzian form. In the strong coupling case depicted in Fig. 1(a-iii), all of the rotational peaks broaden and merge into a single peak. In such a case, because the rotational energy levels are mixed, we can adopt the angular coordinate representation to describe the rotational dynamics. Under the high-temperature approximation without a rotational potential, the QFPE in \eqref{FPEQ} coincides with the CFPE. For this reason, our quantum results exhibit absorption profiles that are similar to those in the classical case. This does not mean, however, that the quantum results approach the classical results in the strong damping case, because we always have low temperature quantum correction terms in Eqs.\eqref{eq:QMEiso}, \eqref{eq:RRE}, \eqref{eq:RFPE} and \eqref{FPEQ} in the low temperature case, where quantum effects play a significant role,
as illustrated in Refs. \onlinecite{Tanimura-2014,Tanimura-2015, SakuraiJPC11, KatoJPCB13}.

In the very strong coupling (overdamped) case depicted in Fig. 1(a-iv), the difference between the quantum RISB results and the other results becomes large again, because our perturbative treatment of the quantum RISB calculation based on the eigen-state representation of the system becomes inappropriate: In such a case, the energy states of the system become continuous, because the states of the system and bath are entangled due to the strong system-bath interaction.
The coordinate representation of the equation of motion Eq. \eqref{FPEQ}, however, can be used even in the overdamped case, due to the fact that the system energy described in coordinate space is continuous, although we have to include low temperature quantum correction terms in order to obtain an accurate spectrum.\cite{Tanimura-2015} 
Then, from the similarity of the quantum and classical RCL results described by Eqs \eqref{FPEQ}, we infer that the quantum RISB results should be similar to the quantum RCL results appearing in Fig. 1(a-iv) in the strong coupling regime if we can accurately solve the RISB model quantum mechanically. However, this must be confirmed by computing spectra using both models at low temperature, where quantum effects play a significant role.

In the high temperature case depicted in Fig. 1 (b-i)-(b-iv), the RCL results are qualitatively similar to the CLE results at any coupling strength, while we observe rotational bands in the weak coupling case in the RISB result. 
This is because the high temperature limit ($\beta \rightarrow 0$) is effectively the same as the classical limit ($\hbar \rightarrow 0$) for a harmonic heat bath, as can be seen in the QME approach, in which the temperature appears as $\beta\hbar$.  While the spectrum exhibits a Gaussian-like profile in Fig.1 (b-ii), as described by Eq. \eqref{eq:Gaussian}, it becomes a Lorentzian-like profile in Fig.1 (b-iv), as described by Eq. \eqref{eq:Lorentzian}. Because the Kubo oscillator theory is a classical theory, this narrowing behavior of the spectrum is regarded as having a classical origin.

 We are able to simulate these phenomena from the quantum regime to the classical regime uniformly because our RISB model has a proper classical limit, although there is a discrepancy in Fig. 1(a-iv) due to the perturbative treatment of the QME approach.
We should note that if we use the rotating wave approximation to derive the QME, we cannot account for the transition from the Gaussian-like to Lorentzian-like spectral profile due to the improper treatment of the thermal activation processes, while the positivity of the reduced density matrix is maintained.\cite{Tanimura-2015} 

\begin{figure}
\centering
\includegraphics[width=0.9\textwidth]{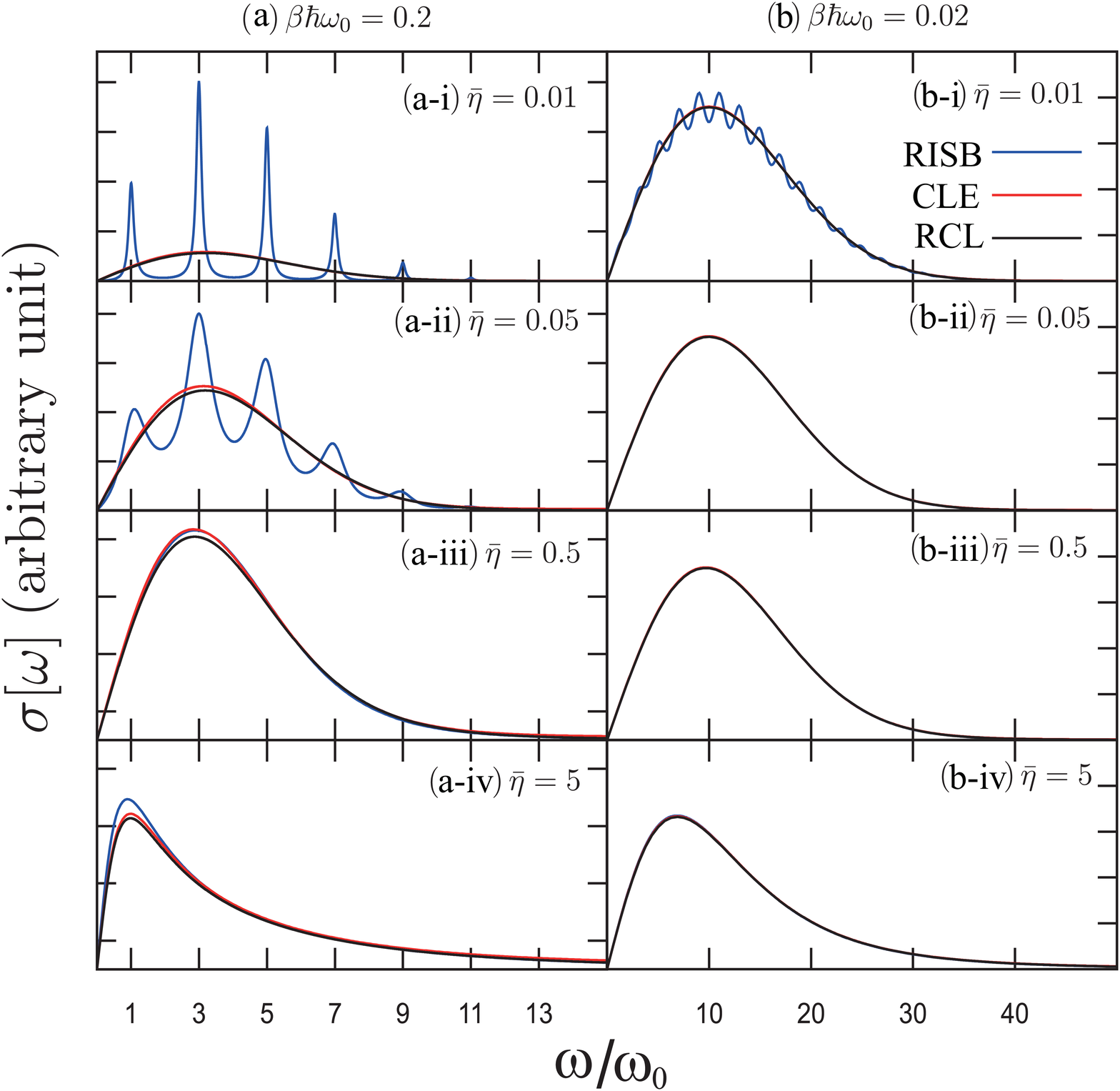}
\caption{
Rotational absorption spectra, $\sigma[\omega ]$, in (a) the moderate-temperature case ($\beta \hbar \omega_0 =0.2$) and (b) the high-temperature case ($\beta \hbar \omega_0 =0.02$) for four values of the coupling strength, $\bar{\eta} = \eta / \hbar$: (i) 0.01, (ii) 0.05, (iii), 0.5, (iv) 5. The blue, black, and red curves represent the quantum RISB result, quantum RCL result, and classical Langevin equation (CLE) result, which corresponds to the classical limit of the RISB and RCL results, respectively. The quantum RCL result is calculated from the analytical expression presented in Refs. \onlinecite{Suzuki-2001,Suzuki-2002} }
\end{figure}

\section{Conclusion}
In this work, we introduced the RISB model in order to describe the dynamics of a two-dimensional rigid rotator in a dissipative environment. As we demonstrated, the RISB model allows us to explain the characteristic feature of the rotational spectrum as a function of the system-bath coupling and bath temperature in a unified manner. This characteristic feature is a transition of the peak profiles from  discretized rotational bands to a Lorentzian-like peak through a Gaussian-like peak.
{Here, we calculated absorption spectrum that is described by the correlation function of the cosine function. However, this approach can be extended straightforwardly to calculate rotational Raman spectrum that is described by the correlation function of Legendre polynomial.}

In this paper, we limited our analysis, using the perturbative and Markovian QMB approach for the relatively high temperature cases. As a result, we were not able to obtain an accurate prediction of the motional narrowing peak in the strong coupling case at moderate temperatures. Although it is computationally demanding, we can study the effect of a non-Markovian environment at low temperature using the hierarchical equations of motion (HEOM) approach.\cite{Tanimura-1989, Tanimura-1990,Ishizaki-2005,Tanimura-2006,Tanimura-2014,Tanimura-2015} Because understanding the noise correlation in both isotropic and anisotropic environments is very important for many areas of physics, chemistry and biology, such an extension is necessary.
This formalism in the Wigner representation is ideal for studying rotator systems, because it allows for the treatment of rotationally invariant systems with any potential profiles, in addition to the inclusion of an arbitrary time-dependent external field, utilizing periodic boundary conditions.\cite{TanimuraJCP92,KatoJPCB13}  Moreover, because we can compare quantum results with classical results obtained in the classical limit of the equation of motion for the Wigner distribution, this approach is effective for identifying purely quantum effects.\cite{TanimuraJCP92,KatoJPCB13,SakuraiJPC11}  Superconducting quantum interference devices (SQUIDs) can also be investigated using the same framework.\cite{Caldeira-1987,Ambegaokar} 

The extension of the RISB model from two dimensions to three dimensions is also necessary, because the dynamics of 2D rotators and 3D rotators are different {even in the classical case.\cite{Pierre,Blokhin} In addition, the effect of rotational potential is important in most chemical systems, for example, to analyze the hindered rotation of a molecular system. In the present formalism, rotational potential is easily included in Eqs.\eqref{eq:QMEiso} without increasing computational costs.}
As a future investigation, we plan to extend the present study in such directions.

\section*{Acknowledgments}
Y.~T.~is supported by JSPS KAKENHI Grant Number A26248005.

\appendix

\section{Wigner representation of the QME for the RISB model}
For a system described by an angular coordinate, a discrete Wigner distribution is often employed.\cite{Mukunda,Bizarro}  In the overdamped case, we can employ a regular Wigner distribution even in this periodic case, because the rotational motion relaxes quickly, due to the large viscosity, and hence we can extend the domain of $\theta$ from $- \pi \le \theta < \pi$ to $- \infty < \theta < \infty$. In the Wigner representation, an arbitrary operator $\hat{A}$ is defined as\cite{Wigner-1984,Tanimura-2006, Tanimura-2015, Frensly, TanimuraJCP92, SakuraiJPC11, KatoJPCB13} 
\begin{align}
A_W(p,q) &= \int^\infty_{- \infty} dr e^{- \frac{i p r}{\hbar}} \langle \theta | \hat{A} | \theta'\rangle,
\end{align}
where $q = \frac{\theta + \theta'}{2}$ and $r = \theta - \theta'$. Then, for the density operator $\hat{\rho}$, we have
\begin{align}
W(p,q) = \int^\infty_{- \infty} dr e^{- \frac{i p r}{\hbar}} \langle \theta | \hat{\rho} | \theta'\rangle,
\end{align}
where $W(p,q)$ is the Wigner distribution function.
The kinetic term of the Liouvillian in the Wigner representation is expressed as
\begin{align}
\left( \frac{i \hbar}{2 I} \frac{\partial^2}{\partial \theta^2} - \frac{i \hbar}{2 I} \frac{\partial^2}{\partial \theta'^2} \right) \hat{\rho}(\theta , \theta') \to -\frac{p}{I}\frac{\partial}{\partial q}.
\end{align}
If we assume $\lim_{r \to \pm \infty} \rho(q + \frac{r}{2} , q - \frac{r}{2}) = 0$, the system side of the system-bath interactions is given by
\begin{align}
\int^\infty_{- \infty} dr e^{- \frac{i p r}{\hbar}} \cos(r) \langle \theta | \hat{\rho} | \theta'\rangle &=\frac{1}{2} \int^\infty_{- \infty} dr (e^{- i(\frac{p}{\hbar} + k) r} + e^{- i(\frac{p}{\hbar} - k) r}) \langle \theta | \hat{\rho} | \theta'\rangle \notag \\
&=\frac{W(p + \hbar , q , t) + W(p - \hbar , q , t)}{2}, \notag \\
\int^\infty_{- \infty} dr e^{- \frac{i p r}{\hbar}} \sin r \frac{\partial}{\partial r} \langle \theta | \hat{A} | \theta'\rangle &= - \int^\infty_{- \infty} dr \left( - \frac{i p}{\hbar} \right) e^{- \frac{i p r}{\hbar}} \sin r \langle \theta | \hat{A} | \theta'\rangle -  \int^\infty_{- \infty} dr e^{- \frac{i p r}{\hbar}} \cos r \langle \theta | \hat{A} | \theta'\rangle \notag \\
&=\frac{p}{2 \hbar} (W(p - \hbar  , q , t) -W(p + \hbar  , q , t)  ) -\frac{1}{2} (W(p - \hbar  , q , t) + W(p + \hbar  , q , t)  ). \notag
\end{align}
As a result, the QME in the Wigner representation becomes
\begin{align}
\frac{\partial}{\partial t} W(p,\theta,t) &= -\frac{p}{I} \frac{\partial}{\partial \theta}W(p,\theta,t) + \eta k_B T \frac{W(p + \hbar  , \theta , t) -2 W(p , \theta , t) + W(p - \hbar  , \theta , t)}{\hbar ^2} \notag \\
&+ \frac{\eta}{I} \left \{\frac{p}{2 \hbar } (W(p + \hbar  , \theta , t) -W(p - \hbar  , \theta , t)  ) + \frac{1}{2}(W(p - \hbar  , \theta , t) + W(p + \hbar  , \theta , t)  ) \right \}  \notag \\
&-\frac{\eta}{2 I} (W(p + \hbar  , \theta , t) -2 W(p , \theta , t) + W(p - \hbar  , \theta , t)).
\end{align}
The distribution as a function of the momentum is slowly changed in the high temperature case, and we can approximate the dissipation terms as follows:
\begin{align}
\frac{W(p + \hbar  , \theta , t) -2 W(p , \theta , t) + W(p - \hbar  , \theta , t)}{\hbar ^2} &\approx \frac{\partial^2 W (p,\theta,t)}{\partial p^2}, \notag \\
\frac{W(p + \hbar  , \theta , t) -W(p - \hbar  , \theta , t)}{2 \hbar } &\approx \frac{\partial W(p,\theta,t)}{\partial p}, \notag \\
W(p + \hbar  , \theta , t) + W(p - \hbar  , \theta , t) &\approx 2 W(p  , \theta , t).
\label{eq:approx}
\end{align}
Thus we have
\begin{align}
\frac{\partial}{\partial t} W(p,\theta,t) &= -\frac{p}{I} \frac{\partial}{\partial \theta}W(p,\theta,t) + \frac{\eta}{I} \frac{\partial }{\partial p}\left( p+ \frac{I}{\beta} \frac{\partial}{\partial p} \right) W(p   , \theta , t).
\end{align}
This is the quantum Fokker-Planck equation,\cite{CaldeiraPhysica83,Tanimura-2006} which is identical to the Kramers equation in the classical limit.\cite{Kramers}

\bibliography{aipsamp}

\end{document}